\documentclass[a4paper]{jpconf}
\usepackage{graphicx}
\usepackage{subfig}
\usepackage{tikz}
\usetikzlibrary{shapes.geometric,calc}
\usetikzlibrary{knots}
\usetikzlibrary{arrows.meta,shapes.misc,decorations.pathmorphing,calc}
\usetikzlibrary{decorations.markings}
\usetikzlibrary{arrows.meta,
                quotes
                }
\usetikzlibrary{arrows}

\tikzset{->-/.style={decoration={
  markings,
  mark=at position #1 with {\arrow{>}}},postaction={decorate}}}
\tikzset{-<-/.style={decoration={
  markings,
  mark=at position #1 with {\arrow{<}}},postaction={decorate}}}

\def\d{{\rm d}}
\begin{document}
\title{Magnetic winding – a key to unlocking topological complexity in flux emergence}

\author{D MacTaggart$^1$, C Prior$^2$}

\address{$^1$School of Mathematics and Statistics, University of Glasgow, Glasgow, G12 8QQ, UK}

\address{$^2$Department of Mathematical Sciences, Durham University, Durham, DH1 3LE, UK}

\ead{david.mactaggart@glasgow.ac.uk}

\begin{abstract}
Magnetic helicity is an invariant of ideal magnetohydrodynamics (MHD) that encodes information on the topology of magnetic field lines. It has long been appreciated that magnetic topology is an important constraint for the evolution of magnetic fields in MHD. In applications to the solar atmosphere, understanding magnetic topology is crucial for following the evolution and eruption of magnetic fields. At present, magnetic helicity flux can be measured in solar observations but the interpretation of  results is difficult due to the combination of confounding factors. We propose that a renormalization of helicity flux, the \emph{magnetic winding}, can be used to detect more detailed topological features in magnetic fields and thus provide a more reliable signature for predicting the onset of solar eruptions.
\end{abstract}

\section{Introduction}
The evolution of magnetic fields in magnetohydrodynamics (MHD) is constrained by the underlying topology of magnetic field lines. An invariant of ideal MHD which is related to this underlying field line topology is \emph{magnetic helicity} - a measure of the mean flux-weighted entanglement of the field lines. Moffatt \cite{moffatt1969} first made the connection between the magnetic helicity of closed magnetic fields and topology through the Gauss linking number. A magnetic field with non-zero helicity has its energy bounded from below by this helicity, i.e. the knottedness and linkage of the field lines ensure some energy content to the field. In many solar physics applications, the helicity is essentially conserved, thus preventing the magnetic field relaxing to a current-free state \cite{berger19842}.

In some applications, we have to consider magnetic fields that are not closed, i.e. they have non-trivial components normal to a boundary. For example, in the solar atmosphere, magnetic regions emerge from beneath the solar surface (the photosphere) and then evolve in the atmosphere whilst remaining ``connected'' to the solar surface. To model such fields, the solar surface is treated as a lower boundary, since we have no detailed information about the magnetic field structure beneath the surface. 

These extra (open) boundary conditions present a problem for the classical definition of helicity in closed magnetic fields. Since classical helicity is based on a vector potential, the open boundary conditions mean that this quantity is no longer gauge invariant. To remedy this, a different measure of helicity, known as \emph{relative helicity}, can be used that compares the entanglement of two different magnetic fields with the same boundary conditions \cite{berger1984}. This quantity is gauge invariant for open magnetic fields.

In this paper, we discuss relative helicity flux through a planar boundary and show how it reveals that the \emph{winding} of field lines can be used to describe the underlying topological structure of relative helicity (in a similar way to Gauss linkage for classical helicity). We then present a renormalization of the helicity flux, which we refer to as the \emph{magnetic winding flux}. This quantity depends solely on the geometry of field lines and can detect regions of topological complexity more clearly than the helicity. We finish by arguing that magnetic winding should become a staple calculation, alongside helicity, in the analysis of solar observations since both these quantities together provide much more information about the topological complexity of emerging solar magnetic fields than just the helicity alone. 

\section{Helicity and winding fluxes}
With an eye on applications to solar flux emergence later, we will focus on helicity and winding fluxes through a planar boundary representing the solar surface. In solar observations, we cannot measure the components of the magnetic field in the atmosphere, only at the surface. Therefore, the expressions that we will present can be estimated from solar observations.

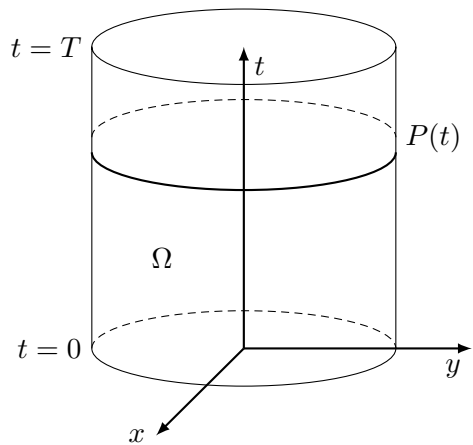
\begin{figure}[h]

  \centering
  \begin{tikzpicture}
\draw[thick,-latex] (0,0,0) -- (3,0,0) node[anchor=north east]{$y$};
\draw[thick,-latex] (0,0,0) -- (0,4,0) node[anchor=north west]{$t$};
\draw[thick,-latex] (0,0,0) -- (0,0,3) node[anchor=east]{$x$};

\draw (-2,4) -- (-2,0) arc (180:360:2cm and 0.5cm) -- (2,4) ++ (-2,0) circle (2cm and 0.5cm);
\draw[densely dashed] (-2,0) arc (180:0:2cm and 0.5cm);

\draw[densely dashed] (-2,2.8) arc (180:0:2cm and 0.5cm);

\draw[thick] (-2,2.6) arc (180:360:2cm and 0.5cm);

\node at (3,2.8) [anchor=east]{$P(t)$};
\node at (-0.8,1.2) [anchor=east]{$\Omega$};
\node at (-2,0) [anchor=east]{$t=0$};
\node at (-2,4) [anchor=east]{$t=T$};
     \end{tikzpicture}  
  
     \caption{Domain of integration $\Omega$. The plane at time $t$ is denoted $P(t)$. The domain is shown as a circular cylinder but this can be generalized to more irregular shapes.}\label{fig_oo}
\end{figure}

All calculations will be performed on a horizontal plane $P$. The magnetic field that intersects with $P$ will change in time due to emergence, submergence, or the horizontal motions due to the foot-points of existing field lines. Therefore, our domain of integration $\Omega$ will consist of ``stacked'' planes $P$ at different times, as indicated in Figure \ref{fig_oo}.

In Figure \ref{fig_oo}, the plane $P$ is shown to be finite but the following results also apply if $P$ has no side boundaries (assuming the magnetic field decays rapidly enough with distance from the plane \cite{berger1984}). We adopt a Cartesian basis $\{\mathbf{e}_1,\mathbf{e}_2,\mathbf{e}_t\}$, where the last vector is in the direction of increasing time. The vector $\mathbf{e}_t$, like $\mathbf{e}_z$, is orthogonal to $P$.

\subsection{Helicity flux}
\begin{figure}
\centering
\begin{tikzpicture}
\draw[thick] (-5,5)--(5,5);
\draw[thick] (-5,0)--(5,0);
\draw[thick] (-5,5)--(-5,0);
\draw[thick] (5,5)--(5,0);

\node at (4, 0.5) [right] {$P$};

\draw [fill] (-1,2.5) circle [radius=.05];
\draw [fill,red] (3.5,4) circle [radius=.05];

\draw[thick,dashed] (-1,2.5)--(3.5,4);

\draw[thick,->] (-1.5,1)--(-1,2.5);
\draw[thick,->] (-1.5,1)--(3.5,4);

\node at (-1.5,1) [below] {$O$};

\draw[thick,dashed] (-1,2.5)--(0.6,2.5);

\node at (-0.05, 2.7) [right] {$\theta$};

\node at (1, 2.2) [right] {$\mathbf{x}$};

\node at (-1.8, 1.8) [right] {$\mathbf{y}$};

\end{tikzpicture}
\caption{\label{angle} Two footpoint intersections with $P$ shown as red and black dots. Their respective position vectors, $\mathbf{x}$ and $\mathbf{y}$, are displayed with reference to a chosen origin $O$. The pairwise winding angle measures the rotation of $\mathbf{x}-\mathbf{y}$ about $O$.}
\end{figure}
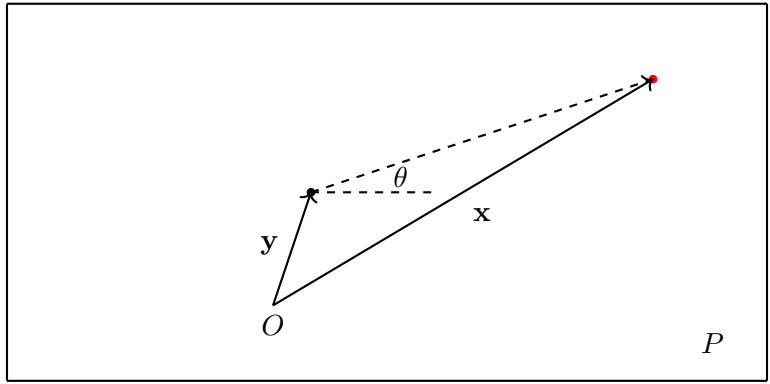

Berger \cite{berger1986} (see also \cite{pariat2005}) showed that the rate of change of relative helicity $H_R$ through $P$ can be written as 
\begin{equation}\label{hel_rate}
\frac{\rm d}{\d t}H_R = -\frac{1}{2\pi}\int_{P\times P}\frac{\d}{\d t}\theta(\mathbf{x},\mathbf{y})B_z(\mathbf{x})B_z(\mathbf{y})\,\d^2x\,\d^2y,
\end{equation}
where $B_z$ is the component of the magnetic field $\mathbf{B}$ orthogonal to $P$, $\mathbf{x}$ and $\mathbf{y}$ are horizontal position vectors on $P$ (marking the intersections of field lines with the plane) and $\theta$ is the the angle of $\mathbf{x}-\mathbf{y}$ on $P$, as indicated in Figure \ref{angle}.

The rate of change of the pairwise winding angle in time is given by
\begin{equation}
\frac{\d}{\d t}\theta(\mathbf{x},\mathbf{y}) = \frac{\d}{\d t}\arctan\left[\frac{(\mathbf{x}-\mathbf{y})\cdot\mathbf{e}_y}{(\mathbf{x}-\mathbf{y})\cdot\mathbf{e}_x}\right] = \mathbf{e}_z\cdot\frac{(\mathbf{x}-\mathbf{y})}{|\mathbf{x}-\mathbf{y}|^2}\times\left(\frac{\d\mathbf{x}}{\d t}-\frac{\d\mathbf{y}}{\d t}\right). 
\end{equation} 
Equation (\ref{hel_rate}) is derived assuming a potential reference field and a particular gauge. We refer the reader to \cite{berger1986} for the details.

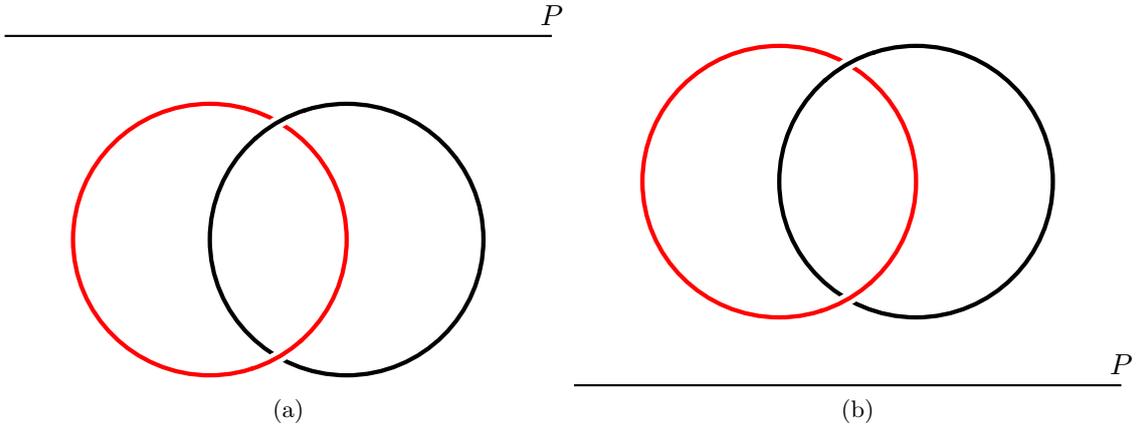
\begin{figure}[h]
\centering
\subfloat[]{\begin{tikzpicture}[scale=0.9]
\draw[thick] (-3,3)--(5,3) node[above]{$P$};
\begin{knot}[
    clip width=3,
    flip crossing=1,
    scale=2
    ]
    \strand [ultra thick, red] (0,0) circle (1.0cm) node [fill=white,inner sep=5pt,yshift=1.5cm](a){};
    \strand [ultra thick, black] (1,0) circle (1.0cm) node [fill=white,inner sep=5pt,yshift=1.5cm](a){};
\end{knot}
\end{tikzpicture}}
\subfloat[]{\begin{tikzpicture}[scale=0.9]
\draw[thick] (-3,-3)--(5,-3) node[above]{$P$};
\begin{knot}[
    clip width=3,
    flip crossing=1,
    scale=2
    ]
    \strand [ultra thick, red] (0,0) circle (1.0cm) node [fill=white,inner sep=5pt,yshift=1.5cm](a){};
    \strand [ultra thick, black] (1,0) circle (1.0cm) node [fill=white,inner sep=5pt,yshift=1.5cm](a){};
\end{knot}
\end{tikzpicture}}
\caption{\label{rings_space} Two closed and linked flux tubes moving through the plane $P$. (a) shows the initial condition at $t=0$ and the fully submerged end state at $t=T_2$. (b) shows the fully emerged state at $t=T_1$.}
\end{figure}

The relative helicity that is determined from equation (\ref{hel_rate}) is only that related to magnetic field passing through $P$ under ideal MHD motion. In order to appreciate how equation (\ref{hel_rate}) measures relative helicity, we now consider a simple thougth experiment. First, consider two thin and closed magnetic flux tubes that are located beneath $P$ and are linked (as shown in Figure \ref{rings_space}(a)). The field lines in each tube are parallel to the tube axis, i.e. we do not need to consider the effects of internal structure on helicity \cite{moffatt1992}. The tubes move up through $P$ until they have completely emerged at $t=T_1$ (Figure \ref{rings_space}(b)). At the later time of $t=T_2$, the tubes have completely submerged again and return to their initial state (Figure \ref{rings_space}(a)). Thus at $t=T_2$, we have, trivially, $H_R=0$ because there is no magnetic field above $P$. From equation (\ref{hel_rate}), we get this result but the calculation has a subtle and important difference. 

\begin{figure}[h]
\centering
\begin{tikzpicture}[scale=0.8]
\draw[thick] (-3,-8)--(5,-8) node[above]{$t=T_2$};
\begin{knot}[
    clip width=3,
    flip crossing=1,
    scale=2
    ]
    \strand [ultra thick, red] (0,0) circle (1.0cm) node [fill=white,inner sep=5pt,yshift=1.5cm](a){};
    \strand [ultra thick, black] (1,0) circle (1.0cm) node [fill=white,inner sep=5pt,yshift=1.5cm](a){};
\end{knot}

\begin{knot}[
    clip width=3,
    flip crossing=1,
    scale=2
    ]
    \strand [ultra thick, black] (1,-2.5) circle (1.0cm) node [fill=white,inner sep=5pt,yshift=1.5cm](a){};
    \strand [ultra thick, red] (0,-2.5) circle (1.0cm) node [fill=white,inner sep=5pt,yshift=1.5cm](a){};
\end{knot}
\end{tikzpicture}
\caption{\label{rings_time} The magnetic field in $\Omega$ at $t=T_2$. Notice the equal and opposite linkage, which translates to equal and opposite winding in equation (\ref{hel_rate}).}
\end{figure}
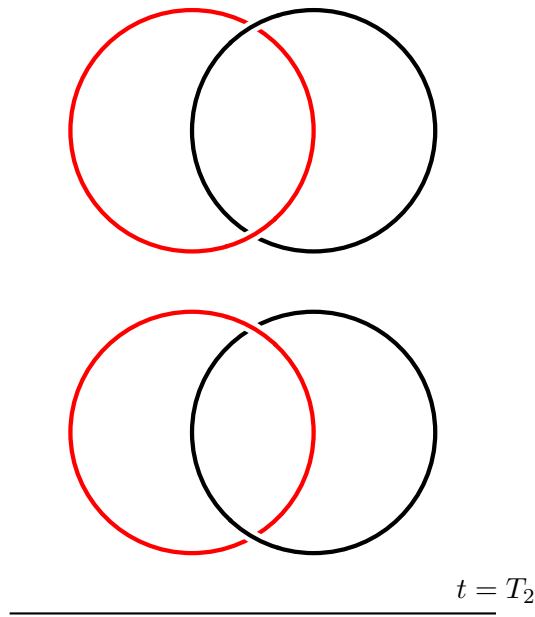

Figure \ref{rings_space} shows the emergence and submergence events at different times in three-dimensional space. This is different, however, from the domain $\Omega$, on which we integrate. Figure \ref{rings_time} shows the field structure in $\Omega$ at $t=T_2$. 
Here we have the full emergence of the tubes (as in Figure \ref{rings_space}(b)) but now there is a copy of the tubes with the opposite sign of linkage. This is because when the tubes submerge through $P$, the points of intersection of the tubes with $P$ rotate in the opposite direction compared to when they were emerging. During emergence, equation (\ref{hel_rate}) records a pairwise winding of magnitude $\theta=2\pi$, due to the fact that the tubes link once and are closed. During subergence, a pairwise winding of $-\theta$ is recorded, i.e. the equal and opposite angle to that recorded in the emergence phase. Hence, since the magnetic flux weighting is constant in time, the result that $H_R=0$ at $t=T_2$ comes from the fact that the recorded topology of the tubes is equal and opposite during the emergence and submergence phases.

Although for this simple example, we could quantify the topology with the Gauss linking number, equation (\ref{hel_rate}) quantifies the topology by measuring the \emph{winding} of field lines passing through $P$. Winding is a more general quantifier of topology as it produces the same result as the Gauss linking number for closed magnetic fields \cite{berger2006} but can also quantify open magnetic fields - in this case, the winding angle $\theta$ would not be an integer multiple of $2\pi$.

As well as measuring the topology of the magnetic field, the value of $H_R$ is also weighted by magnetic flux. Thus, if a magnetic field has weak topology but very strong field strength, the helicity could be large. Also, if the field strength is weak but the topology strong, $H_R$ could also be large. Thus the mixture of information about field line topology and magnetic flux can easily lead to difficulties in interpreting the results of relative helicity flux.  

In the above thought experiment, consider the situation in which the magnetic field experiences some diffusion on $P$ that does not affect the linkage of the tubes. In this situation, $B_z$, and thus the helicity, would decay during the emergence and submergence periods. Therefore, the helicity recorded during emergence would not cancel with that during submergence, as before. This situation would represent a change in relative helicity that does not result from a change in magnetic topology.

\subsection{Winding flux}
One possible solution to the ambiguities of relative helicity, described above, is to renormalize helicity and remove the dependence on magnetic flux. Doing so, we produce a quantity that depends only on the geometry of magnetic field lines. We name this quantity the \emph{magnetic winding} $L$ \cite{berger1986,prior2020}. The flux of $L$ through $P$ is given by 
\begin{equation}\label{wind_flux}
\frac{\rm d}{\d t}L = -\frac{1}{2\pi}\int_{P\times P}\frac{\d}{\d t}\theta(\mathbf{x},\mathbf{y})\sigma(\mathbf{x})\sigma(\mathbf{y})\,\d^2x\,\d^2y,
\end{equation}
where where $\sigma(\mathbf{x})$ is an indicator function marking when the field line at $\mathbf{x}$ moves up or down in $z$, i.e.
\begin{equation}\label{sigma}
\sigma(\mathbf{x}) = \left\{\begin{array}{ccc}
1 & {\rm if} \quad & B_z > 0, \\
-1 & {\rm if} \quad & B_z < 0, \\
0 & {\rm if} \quad & B_z = 0. \end{array}\right.
\end{equation}
It is clear from the replacement of $B_z$ with $\sigma$ in equation (\ref{wind_flux}) that, given two magnetic fields with different field strengths but equal topologies, $L$ would be the same for both fields but $H_R$ would not.
\section{Flux emergence}
\begin{figure}[h]
\centering
\subfloat[]{\includegraphics[width=7cm]{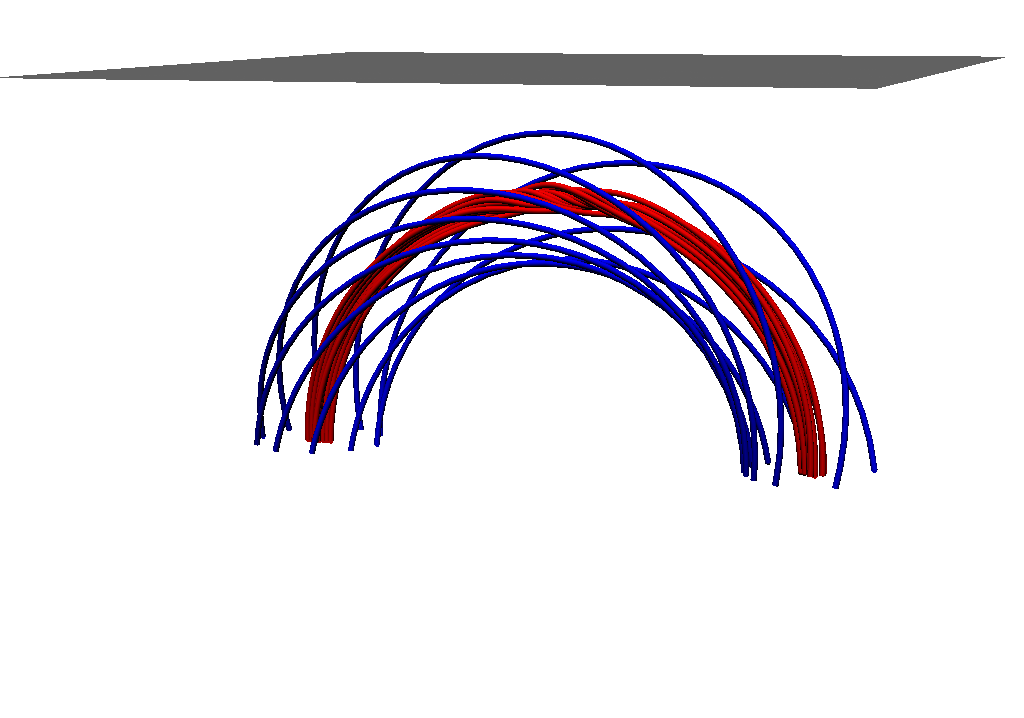}}\subfloat[]{\includegraphics[width=7cm]{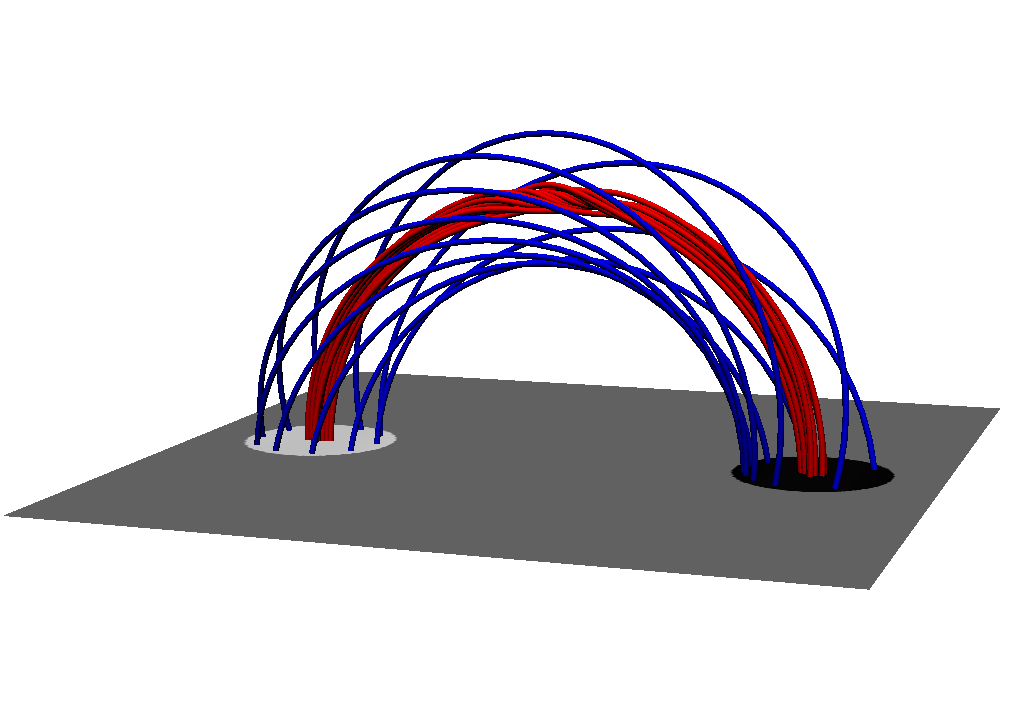}}

\subfloat[]{\includegraphics[width=7cm]{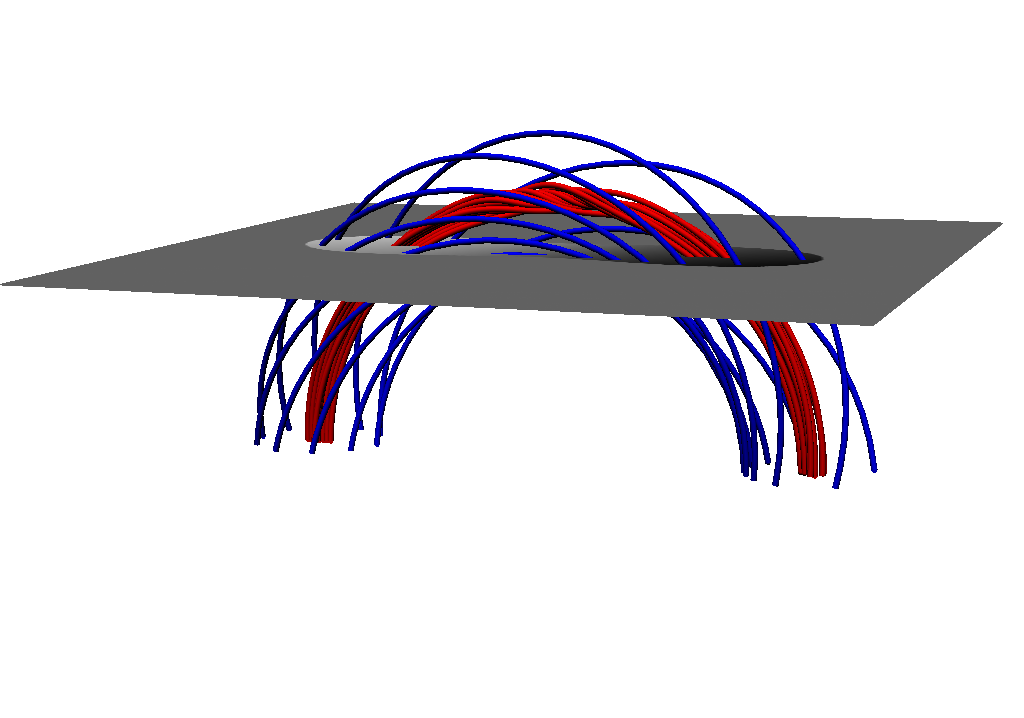}}
\caption{\label{tubes}A magnetic flux tube at different stages of emergence. Details of the magnetic topology are given in the text. (a) $t=0$, pre-emergence; (b) $t=1$, full emergence; (c) $t=1.46$, partial submergence. The plane $P$ is shown and is coloured with the vertical component of the magnetic field, $B_z$. This figure is based on one from \cite{prior2020}.}
\end{figure}

Now that we have expressions for relative helicity and winding fluxes, we can investigate what kind of information each can provide for the emergence or submergence of solar-like magnetic fields. Consider the magnetic flux tube shown in Figure \ref{tubes}. The details of how to construct this tube are given in \cite{prior2020}.

Near the boundary of the flux tube, the magnetic field is twisted, as indicated by the blue field lines. At the centre of the tube, the field lines are weakly-twisted apart from a region of strong twist near the apex, as indicated by the red field lines. The localized region has much stronger twist than the part of the field indicated by the blue field lines. 

\begin{figure}[h]
\centering
\includegraphics[width=9cm]{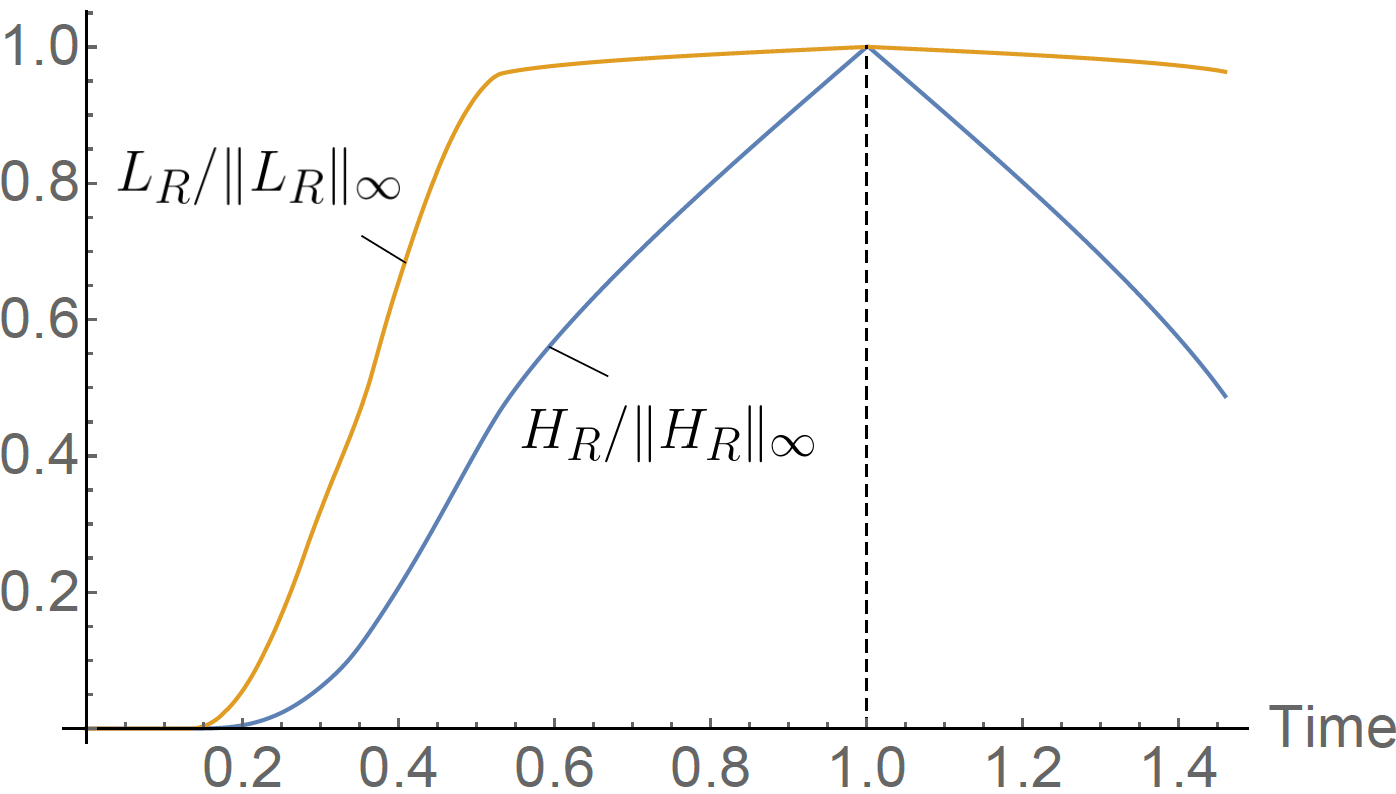}
\caption{\label{rates}The time-integrated relative helicity (blue) and magnetic winding (yellow). Values are normalized with respect to their maximum values. The dashed line indicates when the tube has emerged fully and begins to submerge. This figure is based on one from \cite{prior2020}.}
\end{figure}

In Figure \ref{tubes}(a), at $t=0$, the flux tube is completely beneath the plane $P$. At a later time of $t=1$ (in normalized units), the tube is completely emerged above the plane $P$, as shown in Figure \ref{tubes}(b). At $t=1.46$, the tube is partially submerged with the localized region of strong twist remaining above $P$, as displayed in Figure \ref{tubes}(c). Upon integrating equations (\ref{hel_rate}) and (\ref{wind_flux}), the time-integrated relative helicity and magnetic winding fluxes are displayed in Figure \ref{rates}.

Focussing on the helicity input first, there is an increase during the emergence phase up to $t=1$. In the submergence phase there is a significant removal of relative helicity from above $P$. This behaviour is as expected, but in order to find more detail we need to consider the time-integrated magnetic winding.

For the magnetic winding, the profile shows a much steeper rate of input, compared to the helicity input, during the emergence phase. The difference is due to the fact that helicity is weighted by magnetic flux and winding is not. When the winding picks up the localized region of strong twist passing through $P$, this corresponds to the sharper profile in Figure \ref{rates}. After this region has emerged, the remaining twist in the tube is much weaker and has very little effect on the winding. For the helicity, however, the winding is weighted by magnetic flux which masks the clear distinction between strong and weak twist seen in the winding input.  
 
In the submergence phase, since the localized region of strong twist does not pass through $P$, the effect on the winding is small (it is only picking up the weak twist from the rest of the tube). The helicity, on the other hand, reveals a much stronger decrease since the weak twist is weighted by magnetic flux, which boosts its signal.

This simple toy model shows that the magnetic winding input can detect more localized regions of magnetic topology more clearly than the relative helicity input. This behaviour can also be found in MHD simulations of solar flux emergence. MacTaggart and Prior \cite{prior2019,mactaggart2020} performed detailed analyses of self-consistent MHD simulations of flux emergence, modelling the top of the convection zone through to the corona. Magnetic winding consistently detects the emergence or submergence of localized regions of topological complexity more clearly than the helicity. 

\section{Summary}
In this article, we have discussed how the winding of magnetic field lines acts as the underlying topological description of relative helicity, in a similar way to how Gauss linkage underpins the definition of classical helicity. The relative helicity flux measures this winding weighted by magnetic flux. By removing the magnetic flux from the helicity calculation, we produce a measure of magnetic winding that depends only on the geometry of field lines and not on the strength of the magnetic field. It is shown that the magnetic winding flux can be used to detect specific regions of topological complexity in magnetic fields more clearly than the relative helicity flux.

Observations of relative helicity input into the solar atmosphere currently make use of equation (\ref{hel_rate}). Since equation (\ref{wind_flux}) is very similar in form, it should not be too difficult to also calculate this quantity in observations. Both helicity and winding together would provide a much more complete picture of the topological complexity of emerging solar magnetic fields.

For further reading, a detailed description of theoretical aspects of magnetic winding, and its relation to helicity, can be found in \cite{prior2020}. For the behaviour of helicity and winding in flux emergence, we point the reader to \cite{prior2019,mactaggart2020}.

\section*{References}
\bibliography{iopart-num}

\end{document}